# *Principles in the Evolution of Metabolic Networks*

Hiroki R. Ueda*, John B. Hogenesch†

*Laboratory for Systems Biology, Center for Developmental Biology, RIKEN, 2-2-3 Minatojima-minamimachi, Chuo-ku, Kobe, Hyogo 650-0047, Japan.     †The Genoemics Institute of the Novartis Research Foundation San Diego, CA 92121, USA.

**Correspondence and requests for materials should be addressed to H.R.U. (e-mail: uedah-tky@umin.ac.jp)**

**Understanding design principles of complex cellular organization is one of the major challenges in biology**[1-3]**. Recent analysis of the large-scale cellular organization has revealed the scale-free nature and robustness of metabolic and protein networks**[4-12]**. However, the underlying evolutionary process that creates such a cellular organization is not fully elucidated. To approach this problem, we analyzed the metabolic networks of 126 organisms, whose draft or complete genome sequences have been published. This analysis has revealed that the evolutionary process of metabolic networks follows the same and surprisingly simple principles in Archaea, Bacteria and Eukaryotes; where highly linked metabolites change their chemical links more dynamically than less linked metabolites. Here we demonstrate that this 'rich-travel-more' mechanism rather than the previously proposed 'rich-get-richer' mechanism** [7,13,14] **can generate the observed scale-free organization of metabolic networks. These findings illustrate universal principles in evolution of metabolic networks and suggest marked flexibility of metabolic network throughout evolution.**

To extend earlier observations on the scale-free nature of metabolic networks, we first analyzed the distribution of metabolite links (see Methods for definition) in datasets obtained from 126 organisms; whose draft or complete genome sequences have been published. Figure 1 shows that the distribution of metabolite links follows a power-law in which the probability that a metabolite has links to other $k$ metabolites, decays as, $P(k) \propto k^{-r}$ (see also Supplementary Figure 1). This genome-wide metabolic organization is conserved from Archaea to Bacteria to Eukaryotes with the exponent $r$ of the distribution of metabolite link close to 2 in all 126 organisms.

To investigate the evolutionary process that generates this highly conserved organization across kingdoms, we analyzed the systems-level features of link change of individual metabolites. We calculated the transition probability of link change

$T(k_2, k_1)$ by counting the incidence of these changes for individual metabolites from link level ($k_1$) to link level ($k_2$) in Archaea, Bacteria, Eukaryotes and all 126 organisms (Fig. 2a-d). This analysis revealed that the transition probabilities are not random, but are rather dependent on the before-transition link levels ($k_1$). For example, the transition from a higher link level spreads more widely, whereas transition from a lower link level is confined to the lower link levels (Fig. 2a-d).

To extract the quantitative characteristics in the evolutional process of metabolic networks, we analyzed relationships between the before-transition metabolite link level ($k_1$) and the absolute link change, $|\Delta k| \equiv |k_2 - k_1|$. Figure 2e-h shows that the absolute link change is proportional to the before-transition link level in Archaea, Bacteria, Eukaryotes and all 126 organisms. Stated differently, while the absolute change from higher link levels is higher, it is proportional to the initial before-transition value. To confirm this proportionality, we measured an exponent *s* from log-log plot of absolute link change ($|\Delta k|$) against the before-transition link level $k_1$ (i.e. $|\Delta k| \propto k_1^s$) and found it to be close to 1, in Archaea, Bacteria, Eukaryotes and all 126 organisms (Fig. 2e-h). These results suggest that this proportional evolutional process underlies metabolic networks of Archaea, Bacteria and Eukaryotes.

We then tested whether this proportional evolutional process can regenerate the observed power-law organization of metabolic networks, we calculated the stationary link distribution generated by transition probability matrices of Archaea, Bacteria, Eukaryotes and all 126 organisms (Fig. 2a-d) from arbitrary initial link distributions. We found that the simulated link distribution follows a power-law $P(k) \propto k^{-r}$, regardless of the initial link distribution, with an exponent *r* close to 2 in Archaea, Bacteria, Eukaryotes and all 126 organisms (Fig. 3). These results suggest that this proportional evolutional process is conserved from Archaea, Bacteria, to Eukaryotes, and that this universal evolutionary process can regenerate the observed power-law organization of metabolic networks.

To explore the relationship between proportional dynamics and power-law organization of metabolic networks, we modeled this evolutional process. We hypothesized a proportional evolutional process; where the standard deviation of metabolic link change, $\langle |dk| \rangle$, increases in proportional to the before-transition metabolic link level $k$. We also hypothesized that the average metabolic link changes, $\langle dk \rangle$, is zero, indicating that the link level of each metabolite may increase or decrease with no systematic tendency. From these assumptions we can derive a stationary link distribution and find that it exhibits the power-law distribution with an exponent –2, i. e. $P(k) \propto k^{-2}$ (see Supplementary Analysis 2 for details).

To validate this model, we implemented it with parameters estimated from metabolic networks of all 126 organisms (see Supplementary Analysis 3). Figure 4a shows the transition probability matrix representing the model evolutional process (see Supplementary Analysis 4). The modeled transition probability matrix closely resembled the transition probability matrix calculated from metabolic networks of all 126 organisms (Fig. 4a and 2d). Moreover, we demonstrated that the stationary distribution generated by the model transition probability matrix from any initial distribution exhibits the power-law distribution $P(k) \propto k^{-r}$ with an exponent $r$ close to 2 (Fig. 4b and 2h). For comparison, we also implemented the model with other parameter values and found much less agreement with observed distributions (Supplementary Figure 4). Based on these results, we concluded that metabolic networks from Archaea, Bacteria to Eukaryotes are governed by the same proportional evolutional process; where the link of each metabolite is dynamically changed in proportion to its initial link level, and that this proportionality underlying metabolic network evolution has a critical role in generating the scale-free organization of metabolic networks.

Collectively, these results indicate that a locally imposed constraint on link change gives rise to the global, complex, and presumably robust network organization of metabolic networks. Importantly, this constraint is relative and proportional to the number of metabolic links, implying that selective pressure is proportionally exerted on metabolic link change. This proportionality assures the same level of flexibility for highly connected molecules as well as for less connected molecules. For example, top twenty highest connected molecules of H2O, ATP, ADP, Orthophosphate, Pyrophosphate, NAD+, NADH, H+, NADP+, NADPH, CO2, AMP, Coenzyme A, L-Glutamate, NH3, Pyruvate, Acetyl-CoA, 2-Oxoglutarate, Oxygen and Phosphoenolpyruvate show the largest deviations in their link levels throughout evolution. That the local topology surrounding these critical constituents is also dynamically changing highlights the differing strategies used by various organisms in utilizing these components. For instance, Oxygen is one of the highest connected molecules in aerobic organisms (e.g. Oxygen is the $2^{nd}$, $4^{th}$ and $4^{th}$ highest connected molecules in *H. sapiens*, *M. musculus,* and *R. norvegicus*) whereas it is lowly connected molecules in anaerobic organisms (e.g. Oxygen is the $115^{th}$, $131^{st}$, and $191^{st}$ highest connected molecules in anaerobic organisms *M.thermoautotrophicum*, *S. mutans* and *T. maritima*).

Power-law distributions arise in the metabolic networks with proportional dynamics where highly linked metabolites change their chemical links more dynamically than less linked metabolites. This 'rich-travel-more' mechanism generates power-law distributions in the fixed or slowly growing networks, and contrasts sharply with the 'rich-get-richer' mechanism such as preferential attachment that can generate the power-law distributions in rapidly growing networks, in which highly connected nodes have a higher chance than less connected nodes to linking to new nodes.  To distinguish 'rich-travel-more' mechanism from 'rich-get-richer' mechanism, we also implemented the latter mechanism and calculated its transition probability (Supplementary Analysis 6 and 7, and Supplementary Figure 5).  We found that the observed transition probabilities in Figure 2a-d resembles those from 'rich-travel-more' mechanism, not those from 'rich-get-richer' mechanism.  These results suggest that 'rich-travel-more' mechanism rather than the 'rich-get-richer' mechanism may underlie the evolution of metabolic networks.  Interestingly, we have also analyzed the dynamics of gene expression levels, whose distribution follows power-law, and found that the 'rich-travel-more' mechanism also underlies the gene expression dynamics[15], suggesting the universality of 'rich-travel-more' mechanism across different large-scale biological organizations.

Thorough understanding of biological systems depends on the synthesis and analysis of information as well as the development of predictive models that accurately represent biological organization.  Recently, several studies have investigated the global network structure of transcriptional regulation[16,17] the proteome[18], and metabolism[19-21] and have revealed the universality of the scale-free organization and robust behavior of biological networks[7-11].  This present study demonstrates that a 'rich-travel-more' mechanism underlies the scale-free biological organization, complementing previous work on this generic structure of global cellular organization. Because similar 'rich-travel-more' mechanism has been observed in a number of complex systems, including the world-wide-web[5,22], company size[23], cell number[24] and so on, we expect the universality of 'rich-travel-more' mechanism underlying scale-free organization of biological systems including transcriptional networks[8,9], protein-protein networks[10,11] and social networks[6] as well as those of other non-biological systems beyond the quoted examples.  A next challenge is, thus, the further investigation of the universality of 'rich-travel-more' mechanism underlying other scale-free biological organizations.    Further integration of experimental and theoretical studies may help to uncover other underlying principles governing the complex and dynamic systems of life.

# Methods

**Metabolic Network Data.** Metabolic network information of 136 organisms, whose draft or complete genome sequences are publicly available, were downloaded from LIGAND database[20] on May 28$^{th}$, 2003. We dropped 10 organisms among 136 organisms due to the insufficient metabolic network information (organisms with less than 100 annotated enzymes are dropped) and the metabolic network information of remaining 126 organisms were used for further analysis. We extracted the whole annotated enzymes in each organism from downloaded LIGAND database. We then listed up the reactions catalyzed by the whole enzymes existing in each organism. According to Jeong et al[7], we define the metabolite link between two metabolites if there is a reaction from one metabolite to the other metabolite. We calculated links among metabolites based on the whole reactions existing in each organism. We did not distinguish incoming or outgoing links since all reactions in LIGAND database are bi-directional. The calculated metabolic network data are available upon request.

**Calculation of Link Change of Metabolites.** We calculated the absolute link change $|\Delta k| \equiv |k_2 - k_1|$ for individual metabolites from a certain link level ($k_1$) to another link level ($k_2$), across distinct pairs of 126 organisms. We calculated the absolute link change in all distinct pairs of different organisms.

# Acknowledgements
We thank Dr. Shigeo Hayashi, Dr. Sumio Sugano and Dr. Yutaka Suzuki for stimulating discussion, and Dr. Susumu Goto and Dr. Minoru Kanehisa for metabolic network data.

**Figure 1.** Evolutional Conservation in the Large-scale Organization of Metabolic Networks. **a-d**, The distributions of metabolite link in Archaea (**a**), Bacteria (**b**), Eukaryotes (**c**) and all 126 organisms (**d**) exhibit a power-law distribution in which the probability that a metabolite has *k* links to other metabolites, decays as a power law, $P(k) \propto k^{-r}$. Straight line in each panel represents the estimated power-law distribution. Estimated value of exponent *r* is indicated.

**Figure 2**. Characteristics in Evolution of Metabolic Networks. **a-d**, Transition probability $T(k_2, k_1)$, where a metabolite with link level $k_1$ in a certain organisms changes its link level to $k_2$ in other organism, calculated from metabolic network data in Archaea (**a**), Bacteria (**b**), Eukaryotes (**c**) and all 126 organisms (**d**). Colors in descending order from red to yellow to green represent transition probability. Gray indicates the value of zero (the lack of the transition data). **e-h**, Proportionality in Evolution of Metabolic Networks. The absolute link change ($|\Delta k| \equiv |k_2 - k_1|$) is plotted along the before-transition link level $k_1$ in Archaea (**e**), Bacteria (**f**), Eukaryotes (**g**) and all 126 organisms (**h**). Estimated values of exponent *s* from log-log plot of absolute link change against the before-transition link level (i.e. $|\Delta k| \propto k_1^s$) are indicated.

**Figure 3.** 'Rich-travel-more' Mechanism (Proportional Dynamics) Can Generate the Observed Large-scale Organization of Metabolic Networks. **a-d**, The stationary distributions of metabolite link calculated using transition probability matrices in **Fig. 2a-d** from arbitrary initial distribution of metabolite link. The calculated stationary distributions of Archaea (**a**), Bacteria (**b**), Eukaryotes (**c**) and all 126 organisms (**d**) exhibit a power-law distribution in which the probability that a metabolite has *k* links to other metabolites, decays as a power law, $P(k) \propto k^{-r}$. Straight line in each panel represents the estimated power-law distribution. Estimated value of exponent *r* is indicated.

**Figure 4.** Theoretical Model of 'Rich-travel-more' Mechanism (Proportional Dynamics).

**a,** The model transition probability matrix of 'rich-travel-more' mechanism (proportional dynamics). The model transition probability $T(k_2, k_1)$ represents the probability of link change from a certain link level $k_1$ to other link level $k_2$ during unit time interval. Colors in descending order from red to yellow to green represent transition probability.

**b,** The stationary distributions of metabolite link calculated using the model transition probability matrix from arbitrary initial distribution of metabolite link. The stationary distribution of the theoretical model exhibits a power-law distribution in which the probability that a metabolite has *k* link to other metabolites, decays as a power law. Straight line in panel represents the estimated power-law distribution $P(k) \propto k^{-2}$.

**Supplementary Information for Ueda et al.**

**Supplementary Figure 1.** The Large-scale Organization of Metabolic Networks in 126 organisms. **a,** The distribution of metabolite link in all 126 organisms exhibits a power-law distribution $P(k) \propto k^{-r}$. Estimated values of exponent *r* are plotted.

**b,** The individual distributions of metabolite link. A log plot (two left columns) and log-log plot (two right columns) of distribution of metabolite link are demonstrated. Straight line in each log-log plot panel represents the estimated power-law distribution $P(k) \propto k^{-r}$. Estimated value of exponent *r* is indicated.

**c,** The Zipf distribution of metabolite link in all 126 organisms. Log-Log relationship between the link level (x-axis, $k$) and the rank in the link level (y-axis, $R(k)$) are calculated, $R(k) \propto k^{-q}$. Estimated values of exponent *q* are plotted. Zipf distribution can be derived from the power-law distribution with an exponent –2 (See Supplementary Analysis 5 for details).

**d,** The individual Zipf distributions of metabolite link. Log-log plots of the rank against the link level are demonstrated in individual organisms. Straight line in each log-log plot panel represents the estimated distribution $R(k) \propto k^{-q}$. Estimated value of exponent *q* is indicated in each panel.

**Supplementary Figure 2.** The Analysis of Observed Transition Probability Matrices (See also Supplementary Analysis 1).

**a,** The limit of *n*-th power of transition probability matrices $T^*(k_2, k_1)$. Colors in descending order from red to yellow to green represent transition probability. Gray indicates the value of zero (the lack of the transition data).

**b,** The exponential convergence of *n*-th power of transition probability matrices $T^n(k_2, k_1)$ toward the limit transition probability matrices $T^*(k_2, k_1)$. The average of absolute difference between $T^n(k_2, k_1)$ and $T^*(k_2, k_1)$ are plotted against the power *n* of $T^n(k_2, k_1)$. The average of absolute differences $\langle |\Delta T| \rangle \equiv \langle |T^n(k_2, k_1) - T^*(k_2, k_1)| \rangle$ is exponentially decreased along *n* (i.e. $\langle |\Delta T| \rangle \propto e^{-\frac{n}{q}}$). Time constant *q* is indicated in each panel.

**c-f,** The developmental processes of *n*-th power of transition probability matrices $T^n(k_2, k_1)$ of Archaea (**c**), Bacteria (**d**), Eukaryotes (**e**) and all 126 organisms (**f**). The time constant of transition probability matrix (obtained in Supplementary Figure 6b) divided by that of *Homo sapiens* transition probability matrix is indicated at the right bottom. The power *n* of $T^n(k_2, k_1)$ is determined by multiplying *1*, 2, 4, 8 and 32 by this normalized time constant. Colors in descending order from red to yellow to green represent transition probability. Gray indicates the value of zero (the lack of the transition data).

**Supplementary Figure 3.** The Analysis of Model Transition Probability Matrix of 'Rich-travel-more' Mechanism (Proportional Dynamics).

We show *t*-th power of the model transition probability matrix $T^t(k_2, k_1)$, representing the probability of metabolite link change from a certain link level $k_1$ to other link level $k_2$ during time interval $t$. The parameter of the model ($a = 0.65$, $b = 0.05$ and $s = 1$) is estimated from all 126 organisms metabolic network data (Supplementary Analysis 3). For comparison with transition probability matrix of all 126 organisms (Supplementary Figure 2f), we plotted $T^t(k_2, k_1)$ at *t* = 1, 2, 4, 8, 16 and 32.

**Supplementary Figure 4.** The Theoretical Model of 'Rich-travel-more' mechanism (proportional dynamics) with Various Parameters.

We demonstrate the transition probability of the 'rich-travel-more' mechanism (proportional dynamics), representing the probability of metabolite link change from a certain link level $k_1$ to other link level $k_2$ during an unit time interval ($T(k_2, k_1)$, two left columns), and the stationary distribution generated by the transition probability ($P(k)$, two right columns) for various parameters: *s* = 0, *s* = 0.25, *s* = 0.5, *s* = 0.75, *s* = 1.25, *s* = 1.5 (**a**), *a* = 0.001, *a* = 0.003, *a* = 0.01, *a* = 0.03, *a* = 0.1, *a* = 0.3 (**b**), *b* = 0.03, *b* = 0.1, *b* = 0.3, *b* = 1, *b* = 3, *b* = 10 (**c**), The default parameters are *s* = 1.0, *a* = 0.65 and *b* = 0.05. Colors in descending order from red to yellow to green represent transition probability. The line in the right panel represents the analytical solution $P(k) \propto (k+b)^{-2s}$.

**Supplementary Figure 5.** The Analysis of Model Transition Probability Matrix of 'Rich-get-richer' Mechanism (Preferential Attachment).

We show *t*-th power of the model transition probability matrix $T^t(k_2, k_1)$, representing the probability of metabolite link change from a certain link level $k_1$ to other link level $k_2$ during time interval $t$. The parameter of the model is $v = \gamma = 0.3$, $b = 0.05$. For comparison with transition probability matrix of all 126 organisms (Supplementary Figure 2f), we plotted $T^t(k_2, k_1)$ at *t* = 1, 2, 4, 8, 16 and 32.

**Supplementary Analysis 1. The Analysis of Model Transition Probability Matrices of 'Rich-travel-more' Mechanism (Proportional Dynamics).**

**Power-law distributions are maintained in the presence of 'rich-travel-more' mechanism (proportional dynamics).**

We first attempted to demonstrate that power-law distributions of metabolite links are 'stable' when subjected to the empirically derived transition probability matrices. We iteratively applied transition probability matrices in Figure 2a-d to the power-law distribution of metabolite links such as those in Figure 1a-d ($P(k) \propto k^{-r}, r \approx 2$). We found that the power-law distributions are invariant when subjected to the transition probability matrices (data not shown). This result indicates that power-law distributions are stationary distribution in the presence of 'rich-travel-more' mechanism (proportional dynamics). We also noted that power-law distributions were preserved and stable although links of individual metabolites are dynamically changed.

**Power-law distributions are generated in the presence of 'rich-travel-more' mechanism (proportional dynamics).**

We next attempted to demonstrate that any initial distribution can be transformed into power-law distributions when subjected to the empirically derived transition over time. We iteratively applied transition probability matrices in Figure 2a-d to arbitrary initial distributions to get the stationary distribution. Regardless of initial distributions, we get the power-law distribution $P(k) \propto k^{-r}$ with an exponent $r$ close to 2 as the stationary distribution (Figure 3a-d). To further explore this finding, we calculated and analyzed the *n*-th power of the transition probability matrix, $T^n(k_2, k_1)$, which represents the probability of metabolite link change from a certain link level $k_1$ to other link level $k_2$ during *n* unit times. Despite the apparent differences in transition probability matrices $T(k_2, k_1)$ of Archaea, Bacteria, Eukaryotes and all 126

organisms(Figure 2a-d), the *n*-th power matrices $T^n(k_2,k_1)$ converge to a similar limit transition probability matrices $T^*(k_2,k_1)$ (Supplementary Figure 2a). Interestingly, the limit matrices $T^*(k_2,k_1)$ of Archaea, Bacteria, Eukaryotes and all 126 organisms are independent from the before-transition link level $k_1$, and each row of $T^*(k_2,k_1)$ follows the power-law $T^*(k_2,k_1) \propto k_2^{-r}$ with an exponent *r* close to 2. This result indicates that the probability of metabolite link changes from any link level to a certain link level $k_2$, and over time follows the power-law, implying that the evolutionary dynamics represented in the transition probability matrices $T(k_2,k_1)$ can reproduce the power-law distributions from arbitrary distributions. In other words, this result indicates that the power-law distribution is generated from in the presence of 'rich-travel-more' mechanism (proportional dynamics).

**Power-law distributions are global attractor in the presence of 'rich-travel-more' mechanism (proportional dynamics)**.

Based on these two results, we conclude that power-law distributions are maintained and generated by the empirically observed 'rich-travel-more' mechanism (proportional dynamics). These results indicate that power-law distributions are global attractors in the presence of 'rich-travel-more' mechanism (proportional dynamics): Any distribution of metabolite links reach power-law distributions after an adequate time, and once distributions of metabolite links reach power-law distributions, these distributions will be preserved.

**Differences observed in transition probability matrices are merely apparent.**

We also demonstrated that differences observed in empirically derived transition matrices are only apparent because different 'snap shots' of single 'rich-travel-more' mechanism (proportional dynamics) can represent the empirically derived transition matrices.

The *n*-th power of the transition probability matrices $T^n(k_2, k_1)$ of Archaea, Bacteria, Eukaryotes and all 126 organisms exponentially converge to the limit transition probability matrices $T^*(k_2, k_1)$ (Supplementary Figure 2b). Towards limit matrices, $T^n(k_2, k_1)$ of Archaea, Bacteria, Eukaryotes and all 126 organisms take the similar developmental processes along with *n* (Supplementary Figure 2c-f). The major difference between these processes is the time constant of their developmental processes (Supplementary Figure 2b). For example, the 3.38-th power of the Archaea matrix corresponds to the 126 organisms' matrix (Supplementary Figure 2c and 2f).

**Hypothesis**

It is worth noting that the above analyses are based on the hypotheses that 'rich-travel-more' mechanism (proportional dynamics) empirically derived from the metabolite link change across different organisms also hold for the metabolite link change along evolution. This was consistent throughout all of our analyses.

## Supplementary Analysis 2. Theoretical Model of 'Rich-travel-more' Mechanism (Proportional Dynamics)

We modeled the evolutional dynamics of metabolic networks as a continuous Markov process (also referred to as a "diffusion" process). We hypothesized that the average of instantaneous link changes, $\langle dk \rangle$, is zero because the link level of each metabolite may increase or decrease but it has no systematic tendency. We also hypothesized that the standard deviation of instantaneous link change, $\langle |dk| \rangle$, increases as the before-transition link level $k$. Transition probability density function of this process is characterized by that $\langle dk \rangle = 0$ and $\langle |dk| \rangle = a(k+b)^s$ for $a > 0$, $k \gg b > 0$, $s \approx 1$ where the average of instantaneous link change $\langle dk \rangle$ and the standard deviation of instantaneous link change $\langle |dk| \rangle$ are defined as follows:

$$\langle dk \rangle \equiv \lim_{\Delta t \to 0} \frac{1}{\Delta t} \int T(k+\Delta k, k) \Delta k \, d(\Delta k)$$

$$\langle |dk| \rangle \equiv \sqrt{\lim_{\Delta t \to 0} \frac{1}{\Delta t} \int T(k+\Delta k, k)(\Delta k)^2 \, d(\Delta k)}$$

where $T$ is transition probability density function and $\Delta k$ is an link change during time interval $\Delta t$.

The forward Kolmogorov equation of this "diffusion" process is written as follows (Feller, W. *An Introduction to Probability Theory and Its Applications,* John Wiley & Sons, Inc., New York, 1967):

$$\frac{\partial P(k,t)}{\partial t} = -\frac{\partial}{\partial k}\left(\langle dk \rangle P(k,t)\right) + \frac{1}{2}\frac{\partial^2}{\partial k^2}\left(\langle |dk| \rangle^2 P(k,t)\right)$$

where $P(k,t)$ is the probability density that a metabolite has an link level $k$ at a time $t$. Using the condition that $\langle dk \rangle = 0$ and $\langle |dk| \rangle = a(k+b)^s$, this equation is rewritten as follows:

$$\frac{\partial P(k,t)}{\partial t} = \frac{1}{2}\frac{\partial^2}{\partial k^2}\left(a^2(k+b)^{2s} P(k,t)\right).$$

The stationary distribution $P(k)$ in the stationary condition $\frac{\partial P(k,t)}{\partial t} = 0$ is the solution of the following equation:

$$\frac{d^2}{dk^2}\left(a^2(k+b)^{2s} P(k)\right) = 0.$$

This can be integrated $P(k) = A(k+b)^{-2s}$ for some $A$. In the condition of $k \gg b > 0$ and $s \approx 1$, the stationary distribution of metabolite link follows the power-law distribution with an exponent close to $-2$ (i.e. $P(k) \propto (k+b)^{-2s} \approx k^{-2}$).

**Supplementary Analysis 3.   Estimation of Parameters from 126 organism Metabolic Network Data.**

We estimated three parameters of the dynamics $\langle |dk| \rangle = a(k+b)^s$.  The parameter $s$ is estimated as $s = 1$ from Fig. 2h.  The parameter $b$ is estimated as $b = 0.05$ from the linear-linear plot of absolute link change ($|\Delta k| \equiv |k_2 - k_1|$) against the before-transition link level $k_1$ in all 126 organisms.  Because the parameter $a$ only affects the time constant of convergence (Supplementary Figure 4), we determined the parameter $a$ as $a = 0.65$ such that time constant for time $t$ of the model transition probability matrix is almost equal to the time constant for power $n$ of all 126 organisms transition probability matrix.

**Supplementary Analysis 4. Calculation of Model Transition Probability Matrix of 'Rich-travel-more' Mechanism (Proportional Dynamics)**

To calculate the model transition probability matrix of 'rich-travel-more' mechanism (proportional dynamics), we first descritize the above partial differential equation (Supplementary Analysis 2) by $\partial k \to 1$ and $\partial t \to 0.0005$, and represent it in the form of transition probability matrix (we call this matrix as the 'basic' matrix). Then, we calculate the 2000th power of this 'basic' matrix, which represents the probability of metabolite link change from a certain link level $k_1$ to other link level $k_2$ during a unit time interval ($\Delta t = 0.0005 \times 2000 = 1$). We define the 2000th power of this 'basic' matrix as the transition probability matrix, $T(k_2, k_1)$ (Fig. 4a). The *t*-th power of the model transition probability matrix, $T^t(k_2, k_1)$, takes the developmental processes similar to all 126 organisms matrix (Supplementary Figure 3).

**Supplementary Analysis 5.  Derivation of Zipf Distribution from Power-law Distribution with an Exponent -2**

If the number of metabolites is $N$, the maximum value of the rank in the link levels is $N$ and the minimum value of the rank in the link level is 1. We also note that $\frac{R}{N}$ is the cumulative probability of link level that is cumulated from highest link level (Using $1 \leq R \leq N$, we can easily derive the important property of the cumulative probability, $0 < \frac{R}{N} \leq 1$). Thus, we can formulate as follows:

$$\frac{R}{N} = \int_{\infty}^{k} P(k')d(-k').$$

We calculate this equation using the power-law distribution with an exponent $-2$ ($P(k) = Ak^{-2}$).

$$\frac{R}{N} = \int_{\infty}^{k} P(k')d(-k') = -\int_{\infty}^{k} Ak'^{-2}\, dk' = -\left[-Ak'^{-1}\right]_{\infty}^{k} = Ak^{-1}.$$

Thus, $R$ is written as follows:

$$R(k) = NAk^{-1} = Bk^{-1} \quad \text{where} \quad B \equiv NA.$$

## Supplementary Analysis 6.   Theoretical Models of 'Rich-get-richer' Mechanism (Preferential Attachment)

We modeled the previously proposed 'rich-get-richer' mechanism (preferential attachment) in the form of forward Kolmogorov equation.   The evolution of $P(k,t)$ is given by the following forward Kolmogorov equation, modified to accommodate the appearance of new metabolites (network growth):

$$\frac{\partial P(k,t)}{\partial t} = -\frac{\partial}{\partial k}\left(\langle dk \rangle P(k,t)\right) - \frac{1}{N(t)}\frac{dN(t)}{dt}P(k,t)$$

where $-\frac{1}{N(t)}\frac{dN(t)}{dt}P(k,t)$ reflects the growth of metabolic networks.   Because of the 'rich-get-richer' mechanism (or preferential attachment), in which highly connected nodes have a higher chance than less connected nodes to linking to new nodes, we hypothesized that the average of instantaneous link changes $\langle dk \rangle$ increases as the before-transition link level $k$: $\langle dk \rangle = \gamma(t)k$ where $\gamma(t)$ represents the link growth rate.

### 'Rich-get-richer' Model I: Barabasi and Albert Model

In the preferential attachment model by Barabasi and Albert (Science. 1999 Oct 15;286(5439):509-512.), both the network growth rate ($v(t) \equiv \frac{1}{N(t)}\frac{dN(t)}{dt}$) and link growth rate ($\gamma(t)$) are explicitly dependent on time:

$$v(t) \approx \frac{1}{t}$$

$$\gamma(t) = \frac{1}{2t}$$

The stationary distribution $P(k)$ in the stationary condition $\frac{\partial P(k,t)}{\partial t} = 0$ is the solution of the following equation:

$$0 = -\frac{\partial}{\partial k}\left(\frac{k}{2t}P(k)\right) - \frac{1}{t}P(k)$$

This can be integrated $P(k) = Ak^{-3}$ for some $A$.

**'Rich-get-richer' Model II: Bornholdt and Ebel Model**

Bornholdt and Ebel applied the original Simon's work to the network evolution (Phys Rev E Stat Nonlin Soft Matter Phys. 2001 Sep;64(3 Pt 2):035104).  Their model hypothesized:

(i) With probability $\alpha$ add a new node and attach a link to it from an arbitrarily chosen node.

(ii) Else add one link from an arbitrary node to a node having $k$ links with probability proportional to its link level.

Network growth rate ($v(t)$) and link growth rate ($\gamma(t)$) in their model are also explicitly dependent on time:

$$v(t) = \frac{1}{N(t)} \approx \frac{1}{t}$$

$$\gamma(t) = \frac{1-\alpha}{t}$$

The stationary distribution $P(k)$ in the stationary condition $\frac{\partial P(k,t)}{\partial t} = 0$ is the solution of the following equation:

$$0 = -\frac{\partial}{\partial k}\left(\frac{(1-\alpha)}{t}kP(k)\right) - \frac{1}{t}P(k)$$

This can be integrated $P(k) = Ak^{-(1+\frac{1}{1-\alpha})}$ for some $A$.

**'Rich-get-richer' Model III: Ueda et al Model (1)**

Network growth rate ($v(t)$) and link growth rate ($\gamma(t)$) of previous 'rich-get-richer' models are explicitly dependent on time. However, we can assume that network growth rate and link growth rate are constant over time without loss of generality:

$$v(t) = v$$

$$\gamma(t) = \gamma$$

In this model, total node ($N(t)$) and total link ($K(t) \equiv \sum_i k_i(t)$ for node $i$) are governed by the following equations:

$$\frac{dN(t)}{dt} = vN(t)$$

$$\frac{dK(t)}{dt} = \sum_i \gamma k_i(t) = \gamma K(t)$$

The stationary distribution $P(k)$ of this model in the stationary condition $\frac{\partial P(k,t)}{\partial t} = 0$ is the solution of the following equation:

$$0 = -\frac{\partial}{\partial k}\left(\gamma k P(k)\right) - vP(k)$$

This can be integrated $P(k) = Ak^{-(1+\frac{v}{\gamma})}$ for some $A$.

## 'Rich-get-richer' Model IV: Ueda et al Model (2)

We can also generalize the above model to deal with the Generalized Pareto Distribution. We assume that the average of instantaneous link changes $\langle dk \rangle$ is linear to the before-transition link level $k$: $\langle dk \rangle = \gamma(k+b)$.

In this model, total node ($N(t)$) and total link ($K(t) \equiv \sum_i k_i(t)$ for node $i$) are governed by the following equations:

$$\frac{dN(t)}{dt} = vN(t)$$

$$\frac{dK(t)}{dt} = \sum_i \gamma(k_i(t)+b) = \gamma K(t) + b\sum_i 1 = \gamma K(t) + bN(t)$$

The stationary distribution $P(k)$ of this model in the stationary condition $\frac{\partial P(k,t)}{\partial t} = 0$ is the solution of the following equation:

$$0 = -\frac{\partial}{\partial k}\big(\gamma(k+b)P(k)\big) - vP(k)$$

This can be integrated $P(k) = A(k+b)^{-(1+\frac{v}{\gamma})}$ for some $A$, which is known to be Generalized Pareto Distribution. In the condition of $k \gg b > 0$ and $v \approx \gamma$, the stationary distribution of metabolite link follows the power-law distribution with an exponent close to $-2$ (i.e. $P(k) \propto (k+b)^{-(1+\frac{v}{\gamma})} \approx k^{-2}$).

## Supplementary Analysis 7. Calculation of Model Transition Probability Matrix of 'Rich-get-richer' Mechanism (Preferential Attachment)

To calculate the model transition probability matrix of 'rich-get-richer' mechanism (preferential attachment), we first descritize the above partial differential equation ('Rich-get-richer' Model IV, Supplementary Analysis 6) by $\partial k \to 1$ and $\partial t \to 0.0005$, and represent it in the form of transition probability matrix (we call this matrix as the 'basic' matrix). Then, we calculate the 2000th power of this 'basic' matrix, which represents the probability of metabolite link change from a certain link level $k_1$ to other link level $k_2$ during a unit time interval ($\Delta t = 0.0005 \times 2000 = 1$). Thus, we define the 2000th power of this 'basic' matrix as the transition probability matrix, $T(k_2, k_1)$. The *t*-th power of the transition probability matrix of 'rich-get-richer' mechanism (preferential attachment) $T^t(k_2, k_1)$ takes the developmental processes different from observed matrix or 'rich-travel-more' mechanism (proportional dynamics) (Supplementary Figure 5). We used the following parameter values:

$v = \gamma = 0.3$

$b = 0.0001$

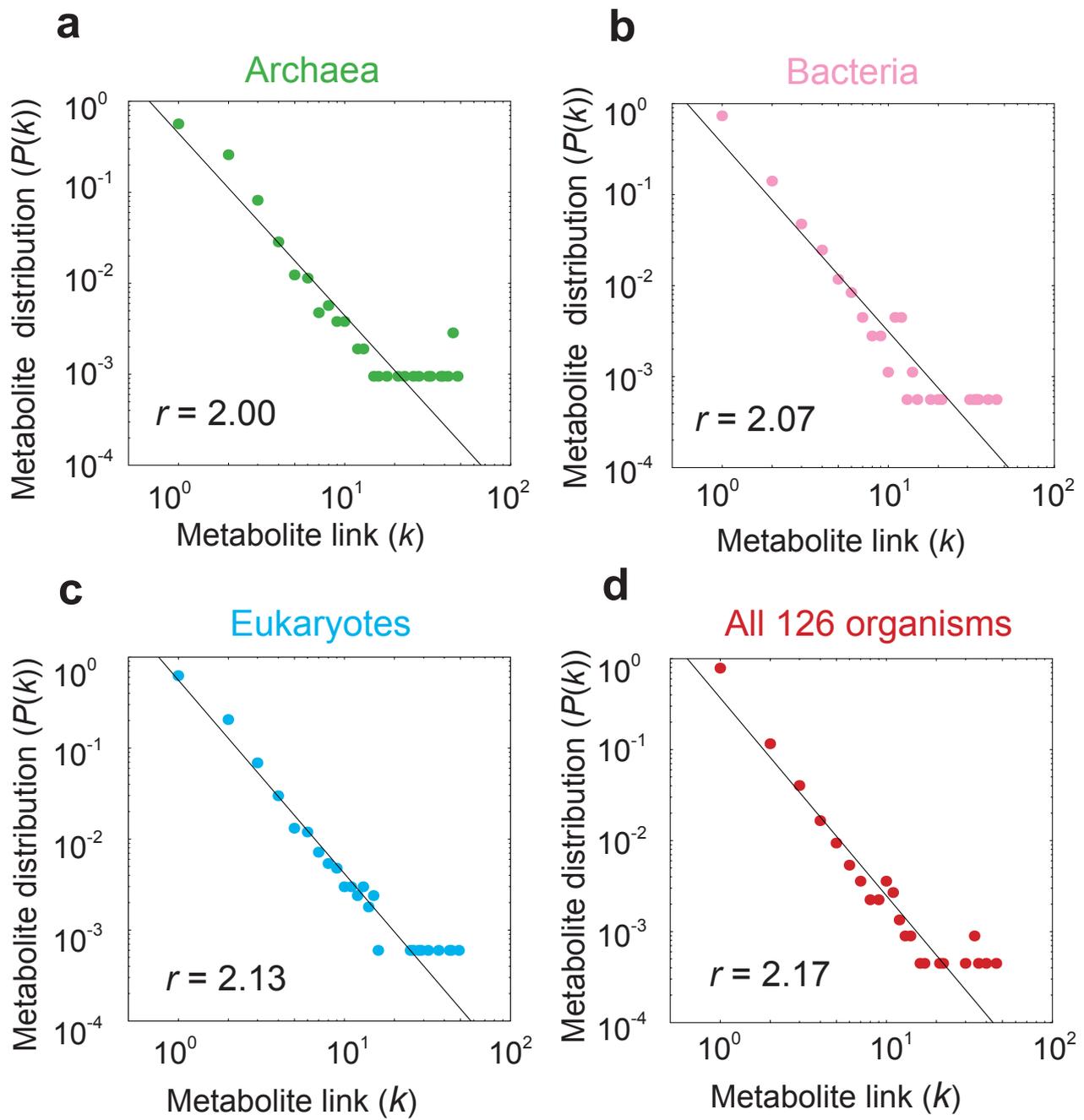

Figure 1 for Ueda et al.

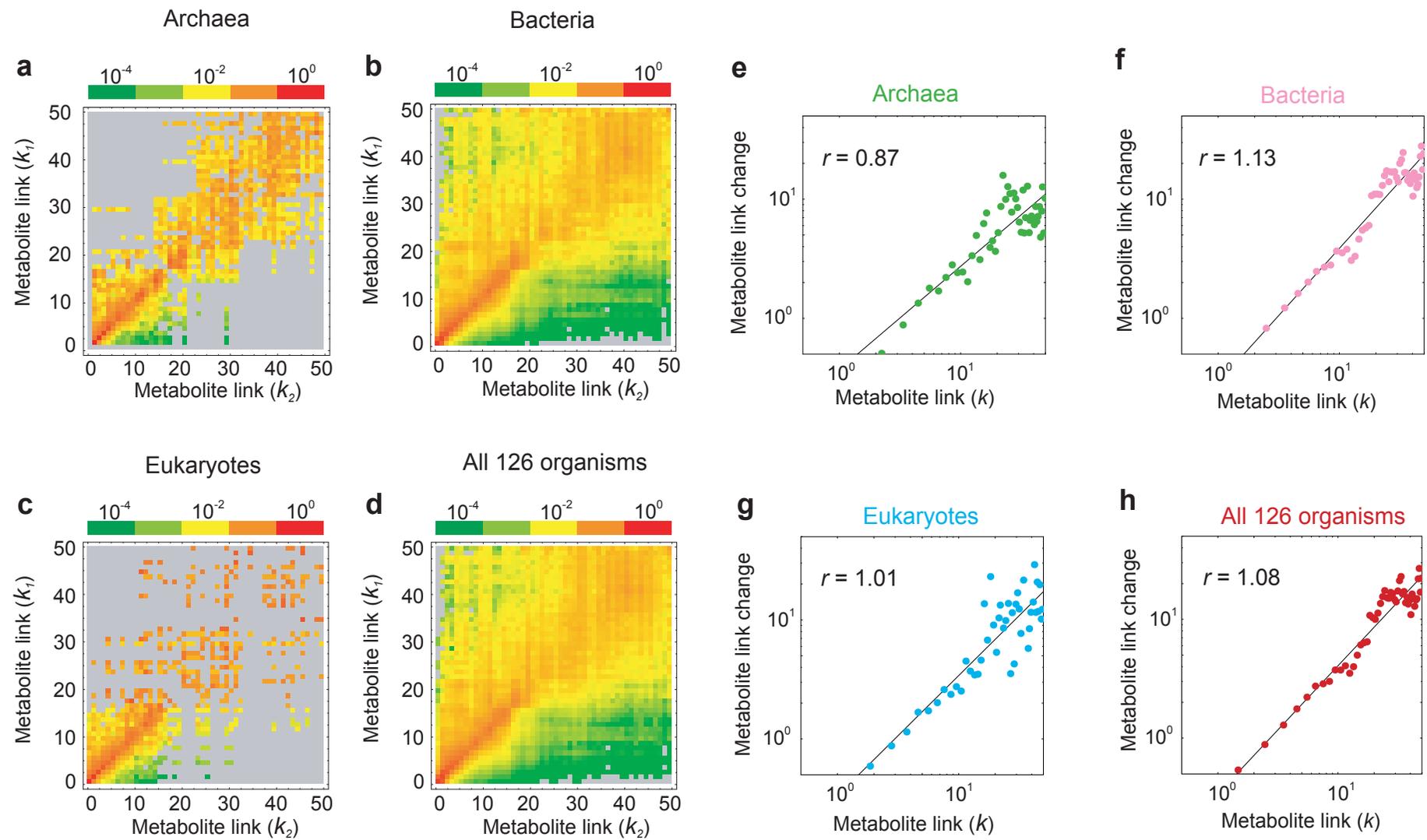

Figure 2 for Ueda et al.

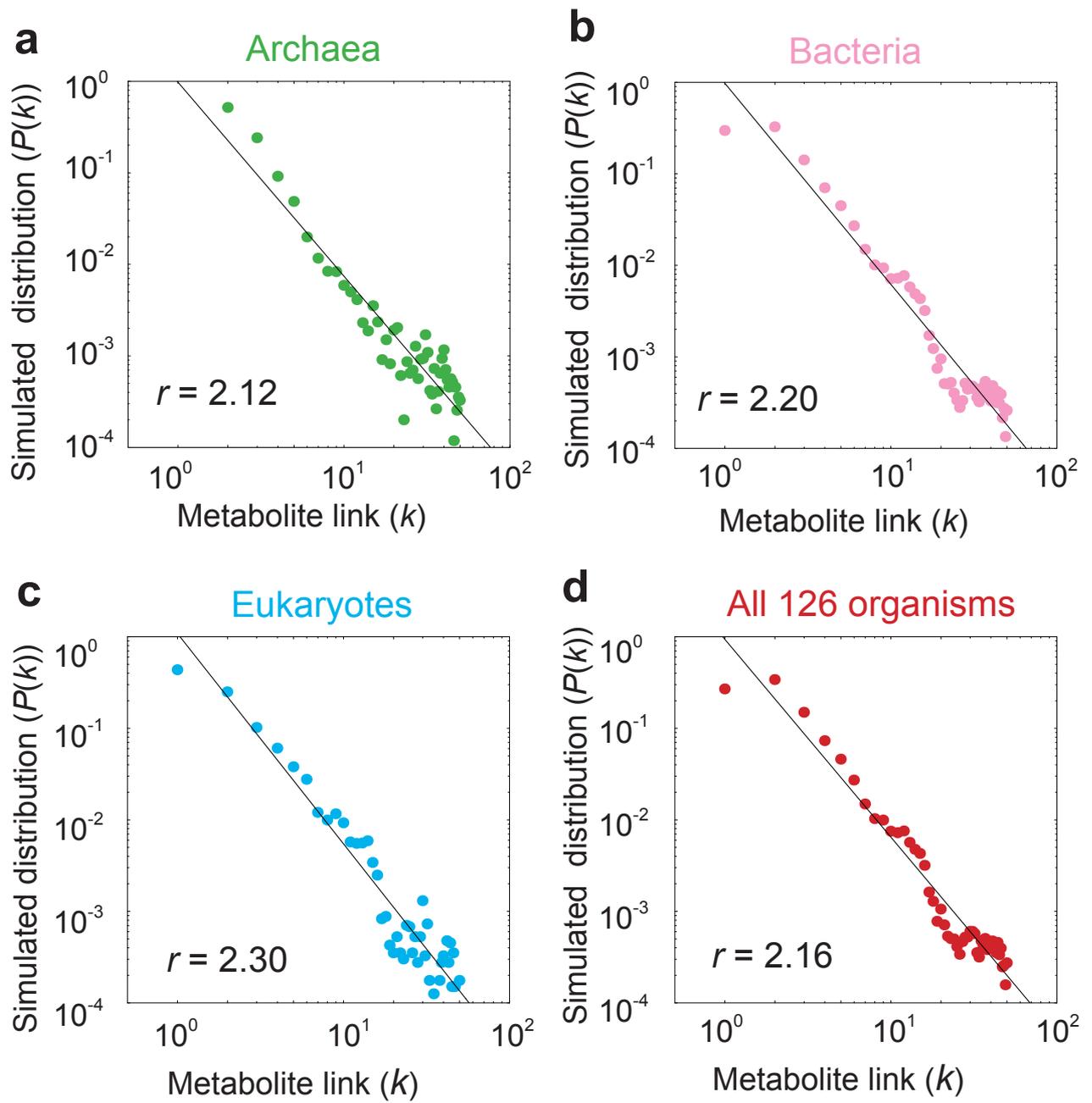

Figure 3 for Ueda et al.

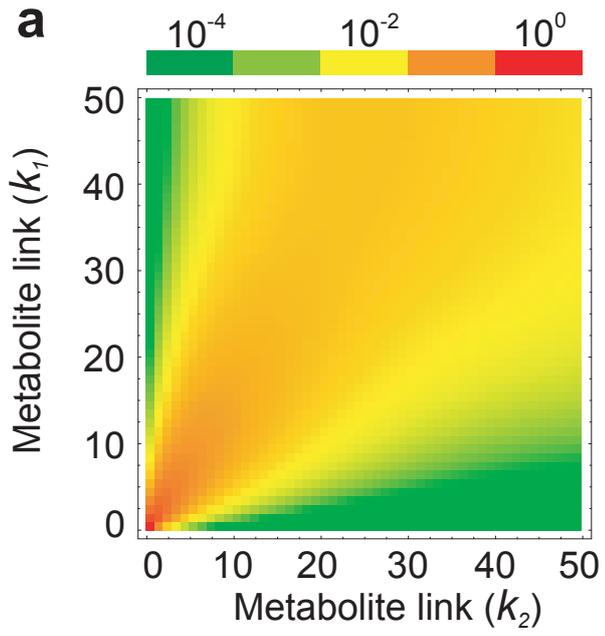 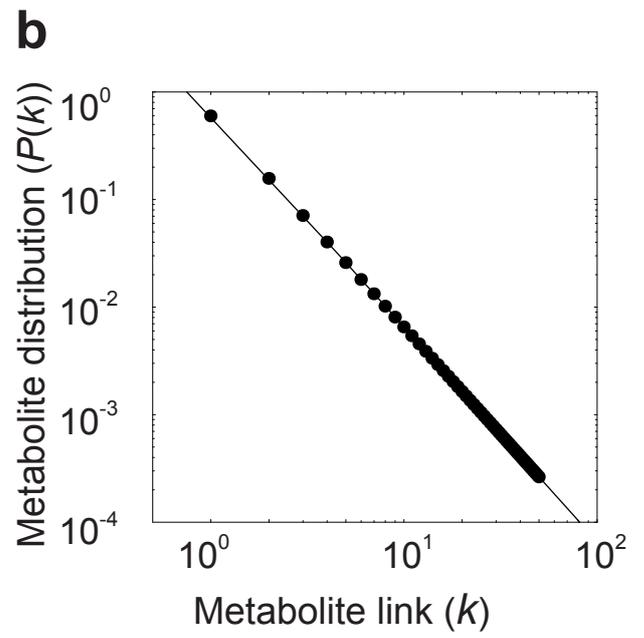

Figure 4 for Ueda et al.